\def\ifmath#1{\relax\ifmmode #1\else $#1$\fi}%

\def\re{\ifmath{{\mathrm{e}}}}

\def\rs{\ifmath{{\mathrm{s}}}}

\documentclass{ws-p8-50x6-00}

\newcommand{\eabe} {\begin{eqnarray}}
\newcommand{\eaen} {\end{eqnarray}}
\newcommand{\eqbe} {\begin{equation}}
\newcommand{\eqen} {\end{equation}}

\newcommand{\anti}[1] {${ \ol \mrm #1 }$}
\newcommand{\pair}[1] {${\mrm {#1 \ol #1} }$}

\newcommand{\mrm}[1] {{\mathrm{#1}}}
\newcommand{\srm}[1] {_{\mathrm{#1}}}
\newcommand{\ol} {\overline}
\newcommand{\CF} {{C\srm F}}
\newcommand{\Nc} {N\srm c}
\newcommand{\Ng} {N\srm g}
\newcommand{\Nq} {N\srm q}

\newcommand{\JHEP}{\it JHEP} 
\newcommand{\EPJC}{{\it Eur.\ Phys.\ J.} C} 
\newcommand{\SJNP}{\it Sov.\ J. Nucl.\ Phys.}
\newcommand{\JETP}{\it Sov.\ Phys.\ JETP} 

\newcommand{\PEfigure}[5]{
\begin{figure}[#1]
\begin{center}
\epsfxsize=#3\textwidth
\epsfbox{#2}
\caption{#4}
\label{f:#5}
\end{center}
\vskip-5mm
\end{figure}
}

\begin{document}

\title{QCD and Multiparticle Production -- The Status of the Perturbative Cascade}

\author{P. Ed\'en}

\address{Nordita, Blegdamsvej 17, DK-2100, Copenhagen, Denmark\\E-mail: eden@nordita.dk}


\maketitle

\abstracts{ 
I discuss recent developments in the QCD cascade
formalism. I focus on the importance of and uncertainties in higher
order corrections to the Modified Leading Log approximation for 
final-state radiation. I also talk about 
the CCFM and LDC evolution equations for
initial-state radiation in DIS.  
}

\section{Introduction}
The QCD cascade is an essential part of our understanding of
multiparticle production in the strong interaction. It is the basis
of many tests of perturbative QCD. Furthermore, in order to
disentangle more and more subtle non-perturbative effects, noting
that non-perturbative QCD and the confinement mechanism is one of the
main unsolved problems of high energy physics, precise knowledge of
the cascade phase is needed.

Thus, there have been activities aiming at improving cascade
calculations,\cite{Dremin,Ochs,jets,Ncomp,CCFM,LDC} and I give a
brief status report of these in this talk. The content may appear
negative and stressing problems, but that is just because I want to
focus on unsolved questions that hopefully will be the topics of
progress in the near future.

\section{$\re^+\re^-$ annihilation}
\PEfigure{t}{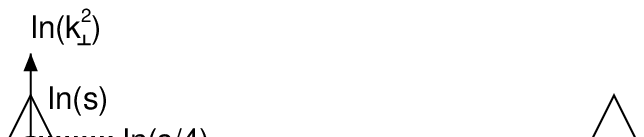}{0.65}{{\bf left:} A triangle in the
(rapidity, $\ln(k_\perp^2)$) plane is a good approximation to the
phase space given by a maximum energy (dashed hyperbola) and a minimum
$k_\perp$ (base line). {\bf right:} After a gluon emission, the total
base line of the two new dipole triangles is larger than $\ln(s)$. The
additional phase space can be drawn as a double sided fold.}{triangle}
It is convenient to illustrate the QCD cascade with the logarithmic
phase space triangle.  With transverse momentum and rapidity expressed
in terms of light-cone momenta,
\eqbe
k_\perp^2 = k_+k_-, \ \ \ y = \frac12\ln(\frac{k_+}{k_-}), 
\eqen
we find that an upper constraint $\sqrt{s}/2$ on the energy
$(k_++k_-)/2$ implies
\eqbe 
|y| \le 
\ln(\frac{\sqrt{s}}{k_\perp})+\ln(\frac{1+\sqrt{1-4k_\perp^2/s}}{2})
\approx \frac12\left[\ln(s)-\ln(k_\perp^2)\right]. \label{e:ymdef}
\eqen 
Thus, the allowed phase space is approximately a triangle in the ($y,
\ln(k_\perp^2)$) plane, cf.\ Fig~\ref{f:triangle}. This triangle is suitable for QCD discussions,
as the gluon emission density off a \pair q dipole behaves like $ \mrm dy\mrm d(\ln(k_\perp^2))$. 

In the dominant case of significantly different $k_\perp$, the emission density for two gluons off a \pair q pair is well approximated by one factor
representing the emission of the first (hardest) gluon $\mrm g$ off the
\pair q colour dipole, times another factor representing the density for the
second (significantly softer) gluon. The latter is given by the summed densities off two
independently emitting dipoles, one for the $\mrm q\mrm g$ dipole,
one for the $\mrm g$\anti q dipole. (There is also a negative term
suppressed by a relative factor $1/\Nc^2$ which reduces the colour
factor in the quark directions from $\Nc/2$ to $\CF$, but this will be
neglected in this simplified discussion.)

The iterative cascade is based on the observation that this factorization generalizes, provided the cascade is strongly ordered in $k_\perp$.
In the parton formulation of QCD cascades,\cite{pQCD} the
emission density off a dipole is split in two terms, representing the
emission off each individual parton of the dipole,
respectively. Colour coherence then leads to the well-known angular
ordering constraint,\cite{ao} which implies that the parton cascade
evolves in angle.

After the emission of the first gluon $\mrm g$, the \pair q phase space triangle is not large enough to represent the two new dipoles. In the right half of
Fig.~\ref{f:triangle}, the additional phase space is drawn as a
double sided fold where the two new triangles meet. The top of this fold
will then have coordinates at 
\eqbe 
y\srm g = \frac12\ln(\frac{s_{\mrm q\mrm g}}{s_{\mrm
g\ol{\mrm q}}}), \ \ \  \ln(k_{\perp \mrm g}^2) = \ln(\frac{s_{\mrm
q\mrm g}s_{\mrm g\ol{\mrm q}}}{s}).  
\eqen 
The fold represents the
phase space for emissions along the new gluon, with rapidities larger
than the latest emission, in other words, the angular ordered cone
around the gluon. For each subsequent emission, a new fold is added, and thus the
triangular picture with folds illustrates the iterative QCD
cascade.

\PEfigure{t}{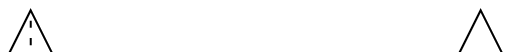}{0.55}{The parton picture of the cascade 
evolves in angle (left) and the dipole picture evolves in 
$k_\perp$ (right).}{evs}

Alternative to the parton formulation, the cascade can be interpreted
 in terms of dipoles, where each gluon emission splits one dipole into
 two new ones.\cite{GGfirst} The dipole cascade evolves in $k_\perp$,
 measured in the rest-frame of the emitting dipole, which makes the
 formalism manifestly Lorenz invariant.  As illustrated in
 Fig~\ref{f:evs}, the parton and dipole formulations of QCD cascades
 are similar, though the evolution parameters differ. Within the
 accuracy of the modified leading log approximation (MLLA),\cite{MLLA}
 they are identical.\cite{GGdipole}

The MLLA of QCD cascades systematically includes corrections
suppressed by a relative factor $\sqrt{\alpha_\rs}$.
It is quite possible to go further and systematically include higher order
corrections embedded in the evolution equations, both in the parton\cite{Capella,Ochs}
and dipole\cite{jets,Ncomp} formalisms. 

The results from such an exercise must be interpreted with some care. 
Corrections beyond MLLA accuracy depend on the treatment of a
very hard first gluon, and also ``moderately ordered emissions'', by
which I mean two subsequent gluon emissions with very similar
$k_\perp$. For these, the factorization
ansatz of the iterative cascade does not
hold. Furthermore, one order beyond MLLA accuracy enter energy
conservation effects. This implies that each emitted gluon not only
adds phase space to the emitting dipole, but also ``eats up'' phase
space in the neighbouring ones. The evolution parameter is then promoted, from a
convenient book-keeping tool of the cascade, to a physical assumption about which gluon indeed is being emitted before others. In a sense,
we then take the classical cascade formulation of quantum mechanical
multiparton production more seriously than we are allowed to.

The study of higher order corrections in cascades is, nevertheless, justified, for a very simple reason.
{\em Corrections one order beyond MLLA are numerically large, for some easily
examined observables like inclusive multiplicity distributions.}

Calculating these corrections, we get an estimate of the important energy
conservation effects and, by comparing different cascade pictures,
we get a hint to the uncertainties in these estimates.

\PEfigure{t}{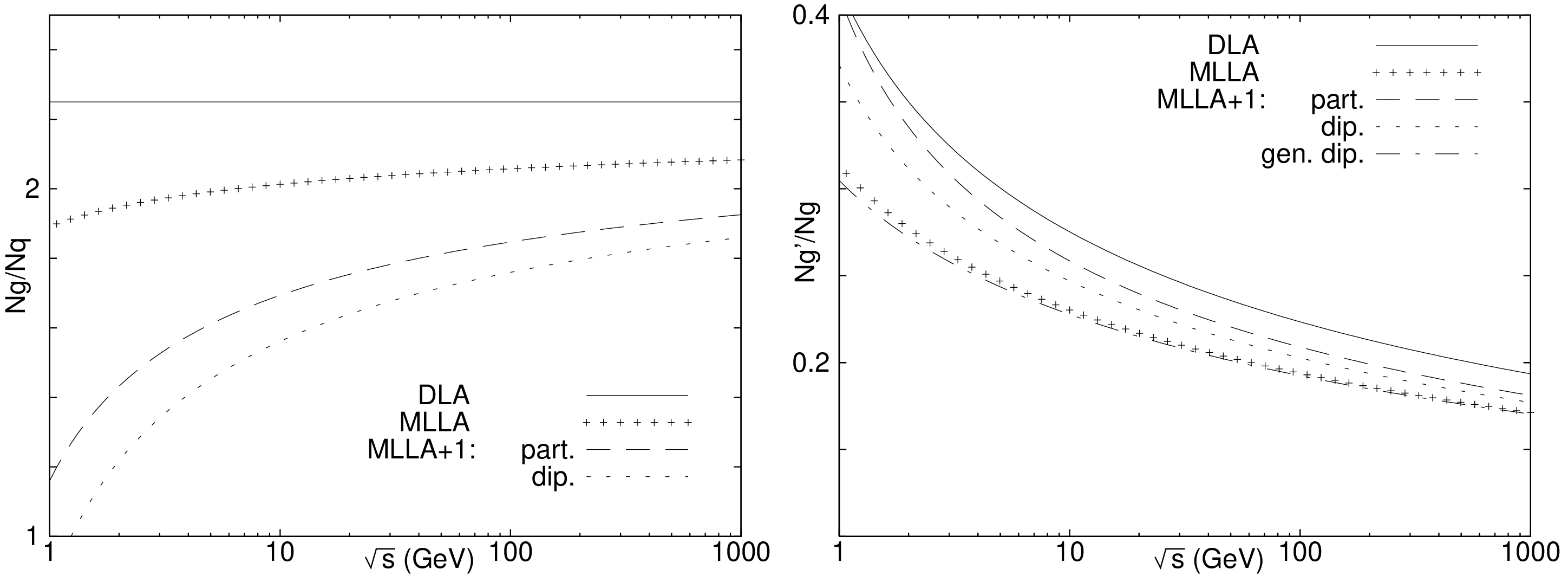}{0.90}{Cascade
predictions for the ratio of multiplicities in gluon and quark jets,
$\Ng/\Nq$ (left) and the anomalous dimension $\Ng'/\Ng$ (right). The
leading order result (double log approximation) is shown with solid
line, MLLA results with crosses. Corrections one order beyond MLLA are
numerically large, and the parton (dashed), dipole (dotted) and
generalized dipole (dash-dotted) equations differ noticeably. The two dipole alternatives give essentially identical predictions for $\Ng/\Nq$, and only the dipole equation results are shown.}{results}

Fig.~\ref{f:results} shows predictions on the ratio of multiplicities in
gluon and quark jets, $\Ng/\Nq$ , and the anomalous dimension
$\Ng'/\Ng$.  Results are obtained with three alternative cascade
evolution equations. One for the parton picture, and two for the
dipole picture. The most recent one, called the ``generalized dipole
evolution equation'' is presented in detail elsewhere.\cite{Ncomp}
The stronger reduction of $\Nq/\Ng$ obtained in the dipole
equation is favoured by data.\cite{Gary,Delphi}

\section{Deep Inelastic Scattering}
\PEfigure{t}{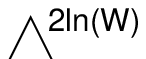}{0.4}{The phase space triangle in
DIS. The lower left corner represents the phase space for final-state 
emissions off the struck quark. The horizontal line represents 
a virtual propagator with $\ln(k_\perp^2) > \ln(k_+k_-)$.}{DIS}

In Deep Inelastic Scattering (DIS), the magnitude of the phase space triangle is determined by the
mass $W$ of the hadronic system. The phase space for final state
emission off the struck quark is given by the photon virtuality scale
$Q$, and is represented by a lower corner of the total triangle. A
virtual propagator, for which $\ln(k_\perp^2) > \ln(k_+k_-)$, can be
represented by a line connecting the $k_+$ and $k_-$ values at the
$k_\perp^2$ height.

In DIS, high-$k_\perp$ emissions, in particular in the target region, are suppressed. Therefore, the magnitude of the triangle is not as directly linked to the multiplicity as in the corresponding case for a \pair q pair. The picture is nevertheless a very useful illustration of DIS events.

\PEfigure{t}{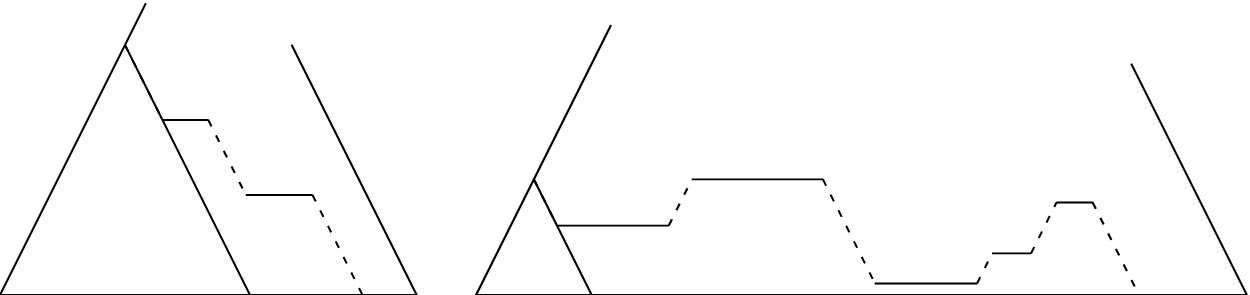}{0.85}{DGLAP (left) evolution dominates for
large $\ln(Q^2)$ and moderate $\ln(1/x)$, while non-DGLAP (right)
evolution is more important for small $\ln(Q^2)$ and large
$\ln(1/x)$. Folds related to emitted partons are not shown.}{DGLAP}
Fig.~\ref{f:DGLAP} illustrates how the cascade for initial-state
radiation in DIS starts at a cut-off virtuality $Q_0$ and an energy
fraction $x_0$ (selected from an input structure function) and then
evolves to the photon interaction point (top of photon triangle).  For
large $Q^2/Q_0^2$ and moderate $1/x$, we have DGLAP evolution, where
emissions ordered in rapidity are strongly ordered in
$k_\perp$.\cite{DGLAP} For small $Q^2/Q_0^2$ and large $1/x$, this
need not be the case.

For high enough energies, the  BFKL evolution,\cite{BFKL} where emissions are strongly ordered in $\ln(1/x)$ but unordered in $k_\perp$, is expected to dominate. The CCFM\cite{CCFM} evolution, and the alternative formulation called the ``linked dipole chain model'' (LDC)\cite{LDC}, extrapolates smoothly between the DGLAP and BFKL regimes. 

Parton cascades in DIS aim at describing both total cross sections and
final-state distributions.  In the following, I will reduce the final-state 
distributions to one observable, which is transverse energy in
the target direction, called ``forward $E_\perp$''.

\PEfigure{t}{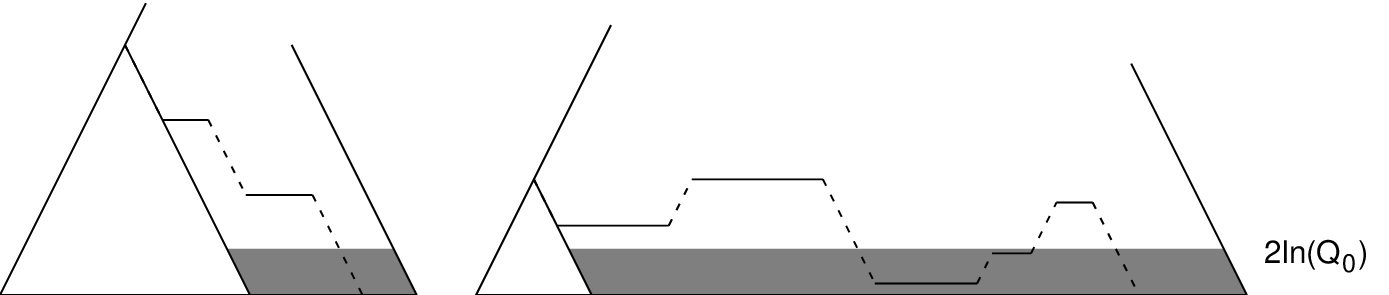}{0.85}{As Fig.~\protect{\ref{f:DGLAP}},
but with a higher value of the cut-off scale $Q_0$, which implies that
perturbative calculations are not trusted in the gray region. In the
non-DGLAP case (right), this gives events whose contribution to
forward $E_\perp$ is not easily included in the formalism.}{scale}

Predictions are obtained from evolution equations convoluted with input structure functions at some cut-off scale $Q_0$.
Fig.~\ref{f:scale} shows what happens when $Q_0$ is raised. The DGLAP evolution predicts no significant forward $E_\perp$, and this simple prediction is of course independent of $Q_0$. A change in $Q_0$ is compensated by a change in the input parton distributions, leaving the total cross section unchanged. 

Non-DGLAP evolution predicts more forward $E_\perp$, but a quantitative prediction may be difficult to make comfortably insensitive to the cut-off scale $Q_0$. With a rise in $Q_0$, some previously allowed propagators of the evolution may drop below the cut-off, which in principle implies that we no longer trust their behaviour to be determined by perturbative physics. Thus, it is not clear how to properly treat emission chains like the one above the gray region on the right hand side of Fig.~\ref{f:scale}, and thus how to properly include their contribution to forward $E_\perp$.

More investigations are needed to answer whether this contribution is numerically important or not. It may be that the discussion above is a problem in principle only, and not in practice, but it is interesting to note that Monte Carlo investigations of the CCFM and LDC schemes find forward $E_\perp$ to be highly sensitive to subleading corrections in the evolution, and that these corrections move predictions too far below data, while leading order calculations agree rather well.\cite{LeifHannes} 

\section{Summary}

As a natural consequence of the success of the parton cascade formalism, higher order corrections are being examined. These are found to be
\begin{itemize}
\itemsep=-2mm
\item theoretically uncertain,
\item numerically important.
\end{itemize}
This should of course be interpreted like this: QCD in Multiparticle
production is an interesting field of physics, suitable for
cooperation between experimentalists and theorists!

\section*{Acknowledgments}
I want to thank the organizers of the conference for a very stimulating week, and in particular the conveners of this session for inviting me to give this talk.

\end{document}